\documentclass[
aps,
prc,
superscriptaddress,
nofootinbib,
floatfix]
{revtex4}
\usepackage{graphicx,
longtable,
color,
array,booktabs,
bm}

\begin{document}
\title{$^{113}$Cd $\beta$-decay spectrum and $g_{\rm A}$ quenching using spectral moments}
%
\author{			Joel~Kostensalo}
\affiliation{		Natural Resources Institute Finland, Yliopistokatu 6B, FI-80100 Joensuu, Finland}
\author{        	Eligio~Lisi}
\affiliation{   	Istituto Nazionale di Fisica Nucleare, Sezione di Bari, 
               		Via Orabona 4, 70126 Bari, Italy}
\author{        	Antonio~Marrone}
\affiliation{   	Dipartimento Interateneo di Fisica ``Michelangelo Merlin,'' 
               		Via Amendola 173, 70126 Bari, Italy}%
\affiliation{   	Istituto Nazionale di Fisica Nucleare, Sezione di Bari, 
               		Via Orabona 4, 70126 Bari, Italy}

\author{        	Jouni~Suhonen}	
\affiliation{		Department of Physics, University of Jyv\"askyl\"a, P.O. Box 35, FI-40014, Jyv\"askyl\"a, Finland}		
\medskip
\begin{abstract}
We present an alternative analysis of the $^{113}$Cd $\beta$-decay electron energy spectrum in terms of spectral moments $\mu_n$, corresponding to the averaged values of $n^{\rm th}$ powers of the $\beta$ particle energy. The zeroth moment $\mu_0$ is related to the decay rate, while higher moments $\mu_n$ are related to the spectrum shape. The here advocated spectral-moment method (SMM) allows for a complementary understanding of previous results, 
obtained using the so-called spectrum-shape method (SSM) and its revised version, in terms of two free parameters: $r=g_{\rm A}/g_{\rm V}$ (the ratio of axial-vector 
to vector couplings) and $s$ (the small vector-like relativistic nuclear matrix element, $s$-NME). We present numerical results for three different nuclear models with the conserved vector current hypothesis (CVC) assumption of $g_{\rm V}=1$. 
We show that most of the spectral information can be captured by the first few moments which are simple quadratic forms (conic sections) in the 
$(r,\,s)$ plane: an ellipse for $n=0$ and hyperbolae for $n\geq 1$, all being nearly degenerate as a result
of cancellations among nuclear matrix elements. 
The intersections of these curves, as obtained by
equating theoretical and experimental values of $\mu_n$, identify the favored values of $(r,\,s)$ at a glance, 
without performing detailed fits.  
In particular, we find that values around
$r\sim 1$ and $s\sim 1.6$ are consistently favored in each nuclear model, 
confirming the evidence for $g_{\rm A}$ quenching in $^{113}$Cd, and shedding light on the role of the $s$-NME.
We briefly discuss future applications of the SMM to other forbidden $\beta$-decay spectra
sensitive to $g_{\rm A}$.

\end{abstract}
\medskip
\maketitle

\section{Introduction}
\label{Sec:Intro}

The search for the rare process of neutrinoless double beta decay ($0\nu\beta\beta$), as well the study of its implications for the fundamental nature of the neutrino field (Dirac or Majorana), represent a major topic in both particle and nuclear physics 
\cite{Adams:2022jwx,Agostini:2022zub,Cirigliano:2022oqy}. The predicted rate of this decay, as well as of other weak-interaction nuclear processes, depends sensitively on the effective value of the weak axial-vector coupling $g_{\rm A}$ that, in nuclear matter, appears to be generally different \cite{Suhonen:2017krv,Suhonen:2019qcd} from the vacuum value  $g_{\rm A}^\mathrm{vac}=1.276$ 
\cite{UCNA:2010les,Mund:2012fq}. In particular, effective quenching factors $q=g_{\rm A}/g_{\rm A}^\mathrm{vac}<1$ have been discussed in a variety of observed $\beta$ and $\beta\beta$ decays, whose Gamow-Teller (GT) nuclear matrix elements are reduced by factors of $q$ and $q^2$, respectively; see, e.g., \cite{Suhonen:2017krv,Suhonen:2019qcd,Chou:1993zz,Faessler:2007hu,Barea:2013bz,Ejiri:2019lfs}. 

While the theoretical connections among disparate values of $q$ and their physical origin are still subject to research 
\cite{Suhonen:2017krv,Ejiri:2019ezh,Gysbers:2019uyb,Cirigliano:2022rmf}, from a phenomenological viewpoint it makes sense to study observables that appear to be particularly sensitive to possible quenching effects. In this context, highly forbidden non-unique $\beta$-decays offer a very interesting avenue, since both their calculated decay rates and energy-spectrum shapes are found to change very rapidly around quenched values $g_{\rm A} \sim 1$, due to subtle  cancellations among large nuclear matrix elements \cite{Haaranen:2016rzs}.  

For the fourth-forbidden non-unique $\beta$ decay of $^{113}$Cd, detailed experimental data are available for the energy spectrum $S(w_e)$ in terms of the $\beta$ energy $w_e$ \cite{Belli:2007zza,COBRA:2018blx}. The data constrain both the 
overall decay rate (or, analogously, the half-life) and the energy spectrum shape (as a function of $w_e$).  
It is highly nontrivial to match the corresponding theoretical predictions with data, since the values of $g_{\rm A}$ that best fit the decay half-life are not necessarily the same that best fit the decay spectral shape and may be in conflict
\cite{Haaranen:2016rzs,COBRA:2018blx,Haaranen:2017ovc}, 
although both indicate large NME cancellations. These two approaches to 
constraining $g_{\rm A}$ in 
 $^{113}$Cd $\beta$ decay, dubbed as half-life and spectrum-shape methods \cite{Haaranen:2017ovc}, have only recently been reconciled 
 by varying a small parameter multiplied by the vector coupling $g_{\rm V}$, namely, the so-called small relativistic nuclear matrix element $s$-NME \cite{Kostensalo:2020gha} (${}^V\!{\cal{M}}^{(0)}_{431}$ in the notation of Behrens and B\"uhring \cite{Behrens1982}), 
the estimates of which, from the conserved vector current (CVC) hypothesis or based on specific nuclear models, are rather uncertain but
crucial for forbidden decays \cite{Kirsebom:2019tjd,Kumar:2020vxr}. 

In particular, by treating the $s$-NME as a free parameter 
to be determined by data in a revised version of the spectrum-shape method (SSM) \cite{Kostensalo:2020gha}, 
both the $^{113}$Cd half-life and spectrum data in \cite{Kostensalo:2020gha,Dawson:2009ni} have been found to match 
well the theoretical predictions of different models,
with consistent values of the quenching factor and the $s$-NME. 
These nontrivial results, which represent a strong indication for $g_{\rm A}$ quenching in $^{113}$Cd, 
deserve in our opinion further analysis, aiming at a better understanding of the comparison of theory and data,
in light of recent and future investigations of forbidden $\beta$-decay spectra in other nuclides 
\cite{Kumar:2020vxr,Kostensalo:2017jgw,Leder:2022beq}. In particular,
we aim at reducing the relevant spectral information to a relatively small set of quantities or parameters to be studied.    

We start from the basic property that any smooth spectrum $S(w_e)$ can be characterized by (and reconstructed from) the series of its moments $\mu_n$, namely, by the spectrum-averaged values of $w_e^n$ for $n\geq 0$ \cite{Feller91,Shohat70}. 
This approach to spectra allows
for the unification of the half-life method (connected to $\mu_0$) and the SSM
(connected to $\mu_{1,2,3,...}$) in a single ``spectral-moment method (SMM).'' 
We show that a few moments $\mu_n$
can capture with high accuracy the whole spectral information in terms of the
two free parameters, $r=g_{\rm A}/g_{\rm V}$ and $s=s$-NME, where $g_{\rm V}$ is the vector coupling (assumed to be unity as in \cite{Kostensalo:2020gha}). 
Furthermore, since the moments $\mu_n$ are simple quadratic forms in $(r,\,s)$, the information contained in the 
infinite family of spectra $S(w_e\,|\,r,\,s)$ can be eventually discretized, with significant conceptual and numerical advantages.

Concerning experimental data, herein we use the absolute $^{113}$Cd $\beta$-decay spectrum of 
\cite{Belli:2007zza} including the energy calibration and systematics assessment
performed in \cite{Belli:2019bqp}.  
Constraints on $(r,\,s)$, as obtained by comparing theoretical and experimental moments, are interpreted
in terms of intersections of iso-moment curves (an ellipse for $\mu_0$ and hyperbolae for $\mu_{1,2,3,...}$).  
In each of three different nuclear models for the NME, such intersections are closest for rather similar values
of the $(r,\,s)$ parameters. Since most of the relevant features appear to be captured by just the first
few moments, the method can be pragmatically applied to future forbidden $\beta$-decay measurements in different 
nuclei, where the available spectral data might be more limited than for $^{113}$Cd.

Our work is structured as follows. In Sec.~\ref{Sec:Moments} we discuss the spectral-moment method, the adopted notation, and 
the numerical values of the first few moments. 
\textcolor{black}{In Sec.~\ref{Sec:Compare} we 
discuss the implications of equating the theoretical and experimental moments in terms of the $(r,\,s)$ parameters.}
We find evidence for $r\sim 1$, corresponding to a multiplicative 
renormalization of the axial-vector coupling, as well as for  
\textcolor{black}{$|s|\sim 1.6$ (with a preference for positive values of $s$)}, 
corresponding to an additive
contribution to vector-like NME, consistent with earlier findings utilizing a different approach and independent 
data  \cite{Kostensalo:2020gha}. We summarize our results and consider further applications of the SMM in Sec.~\ref{Sec:Summary}.  
\textcolor{black}{Technical aspects about NMEs and quadratic forms are discussed in Appendix~\ref{appA} and \ref{appB}, respectively.}
%

\section{Spectral moments: method, notation and numerical values}
\label{Sec:Moments}

In this Section we define the notation used in the SMM to describe 
the $^{113}$Cd $\beta$-decay spectra (experimental and theoretical) in terms of a truncated set of spectral moments. 
Numerical values for such moments are also derived.

\subsection{$\beta$-decay spectrum notation \label{notation}}

Following \cite{Haaranen:2016rzs,Haaranen:2017ovc,Kostensalo:2020gha}, we introduce a dimensionless energy parameter $w_e$, 
\begin{equation}
\label{we}
w_e=\frac{W_e}{m_e}=1+\frac{T_e}{m_e}\ ,
\end{equation}
where $W_e$ and $T_e$ are, respectively, the total and kinetic energies of the electron with mass $m_e$. 

The energy spectrum $S(w_e)$ is defined as
the fractional number of decays $n_e$ per single nucleus and per unit of time $t$ and of energy $w_e$
\begin{equation}
\label{spectrum}
S(w_e) = \frac{d^2n_e}{ dt\, dw_e}\ ,
\end{equation}
where $w_e \in [1,\,w_0]$, and the endpoint $w_0$ is set by the $Q_\beta$-value of the decay ($w_0=1+Q_\beta/m_e$). 
When needed, experimental ($e$) and theoretical ($t$) spectra are distinguished by superscripts,
\begin{eqnarray}
\label{theoexpt}
S^e &=& S^\mathrm{expt},\\ 
S^t &=& S^\mathrm{theo}\ .
\end{eqnarray}

In order to link our formalism with common nuclear physics notation, we remind that 
the total decay rate $\lambda$ (or, equivalently, the half life $t_{1/2}$) is
obtained by integrating $S(w_e)$ over the interval $[1,\,w_0]$:
\begin{equation}
\label{rate}
\lambda =\int_1^{w_0} S(w_e)\,dw_e = \frac{\ln 2}{t_{1/2}}\ .
\end{equation}
However, we shall not use either $\lambda$ or $t_{1/2}$ hereafter, for the following reason.
 
Due to increasingly high backgrounds at low energy, the experimental spectrum is typically reported above a detector-dependent kinetic energy threshold $T_{\rm thr}>0$ \cite{Belli:2007zza,COBRA:2018blx,Kostensalo:2020gha,Belli:2019bqp}
that defines the $w_e$ threshold as $w_{\rm thr}=1+T_{\rm thr}/m_e>1$. Therefore, the decay half life $t_{1/2}$ can be estimated only by extrapolating the measured $S^e$ spectrum down to $T\to 0$, see e.g.\ \cite{Belli:2007zza}. 
However, any adopted extrapolation function may well be different from 
the computed theoretical spectra in the range $[1,\,w_{\rm thr}]$ below threshold. In order to avoid potential biases,  
we shall thus consistently compare the experimental and theoretical spectra only in the energy range above threshold,  
\begin{equation}
\label{range}
w_e \in [w_{\rm thr},\,w_0]\ .
\end{equation}

\subsection{Spectral moments $\mu_n$}

It is well known from statistics that a smooth spectrum $S(w_e)$, defined over an interval $w_e\in[w_{\rm thr},\,w_0]$,
can be described by a series of moments $\{\mu_n\}_{n\geq 0}$ \cite{Feller91,Schmudgen17}. The zeroth moment, defined as 
\begin{equation}
\label{moment0}
\mu_0 = \int_{w_{\rm thr}}^{w_0}S(w_e)\,dw_e\ ,
\end{equation}
encodes the overall spectrum normalization, while the first and higher moments, defined as 
\begin{equation}
\label{moment}
\mu_n = \frac{\int_{w_{\rm thr}}^{w_0}S(w_e)\,w_e^n\,\,dw_e}{\int_{w_{\rm thr}}^{w_0}S(w_e)\,dw_e}\ (n\geq1)\ ,
\end{equation}
encode the spectrum shape information, via the averaged values of increasingly high powers of the main variable.%
\footnote{The so-called central moments, not used herein, are alternatively defined by averaging the powers $(w_e-\mu_1)^n$ for $n\geq 2$. The second central moment is the variance, the third the skewness, and the fourth the kurtosis. } 

Note that, in our case, $\mu_0$ has the dimension of an inverse time, being defined as the decay rate 
in the interval $[w_{\rm thr},\,w_0]$ above threshold (see also Eq.~(\ref{rate}) for the total rate in $[1,\,w_0]$). All the other moments are instead dimensionless. When needed, moments of theoretical and experimental spectra will be distinguished by superscripts: 
\begin{eqnarray}
S=S^{t} &\to& \mu_n=\mu_n^t\ , \\
S=S^{e} &\to& \mu_n=\mu_n^e\ . 
\end{eqnarray}

There is a vast and interdisciplinary literature on the inverse moment problem, namely, on possible methods to 
reconstruct the original function $S(w_e)$ from a finite number of moments 
$\{\mu_n\}_{n=0,\dots,N}$ 
with some approximations \cite{Akhiezer20,Shohat70,Talenti87}. 
While all methods tend to improve their accuracy
for increasing $N$, some may converge faster or better than others, depending on specific 
features of the function(s) $S(w_e)$. 

We have checked that a simple reconstruction algorithm based on an expansion in Legendre polynomials,
as described in \cite{Askey82}, is sufficient enough to  allow for the reconstruction of the $^{113}$Cd spectra at subpercent level (more accurately than is needed for our purposes) in the entire parameter space relevant for this work, with just $N=6$ moments. 
Representative examples of theoretical spectra reconstructed from a finite set of moments
are shown below in Sec.~\ref{theospectra}.
Such results are consistent with (but more general than) 
the findings of Ref.~\cite{Belli:2007zza}, where the experimental spectrum was well approximated in terms of   
a 6$^{\rm th}$ order polynomial function.  

We shall thus limit ourselves to $N=6$ and consider the truncated set of moments
\begin{equation}
\{\mu_n\}=\mu_0,\,\mu_1,\dots,\,\mu_6\ .
\end{equation} 
Actually, as we shall see in several ways, 
interesting results can be obtained by considering just the first two or three moments
out of the above set.  

\subsection{Experimental spectrum $S^e(w_e)$} 

In this work we consider the experimental spectrum $S_e(w_e)$ of $^{113}$Cd as measured in 
\cite{Belli:2007zza}, after a recalibration of the energy scale and the deconvolution of resolution effects as described in 
\cite{Belli:2019bqp} (see Fig.~29 therein). The experimental threshold $T_{\rm thr}=26$~keV \cite{Belli:2019bqp} and the 
endpoint $Q_\beta=323.83$~keV \cite{AME2016} define the analysis range:
\begin{equation}
\label{exptrange}
w_e \in [w_{\rm thr},\,w_0] = [1.051,\,1.634]\ . 
\end{equation}
In this range, the experiment observed $N_e= 2.222\times 10^6$ events for $N_d = 8.858\times 10^{22}$ decaying nuclei over a data-taking time $t=9.929\times 10^6$~s \cite{Belli:2007zza,Belli:2019bqp}. The corresponding decay rate provides the zeroth moment $\mu^e_0=N_e/(N_d \, t)$:
\begin{equation}
\label{mu0exp}
\mu_0^e = 2.526 \times 10^{-24}\,\mathrm{s}^{-1} \ .
\end{equation}

\begin{figure}[t!]
\begin{minipage}[c]{0.56\textwidth}
\includegraphics[width=0.56\textwidth]{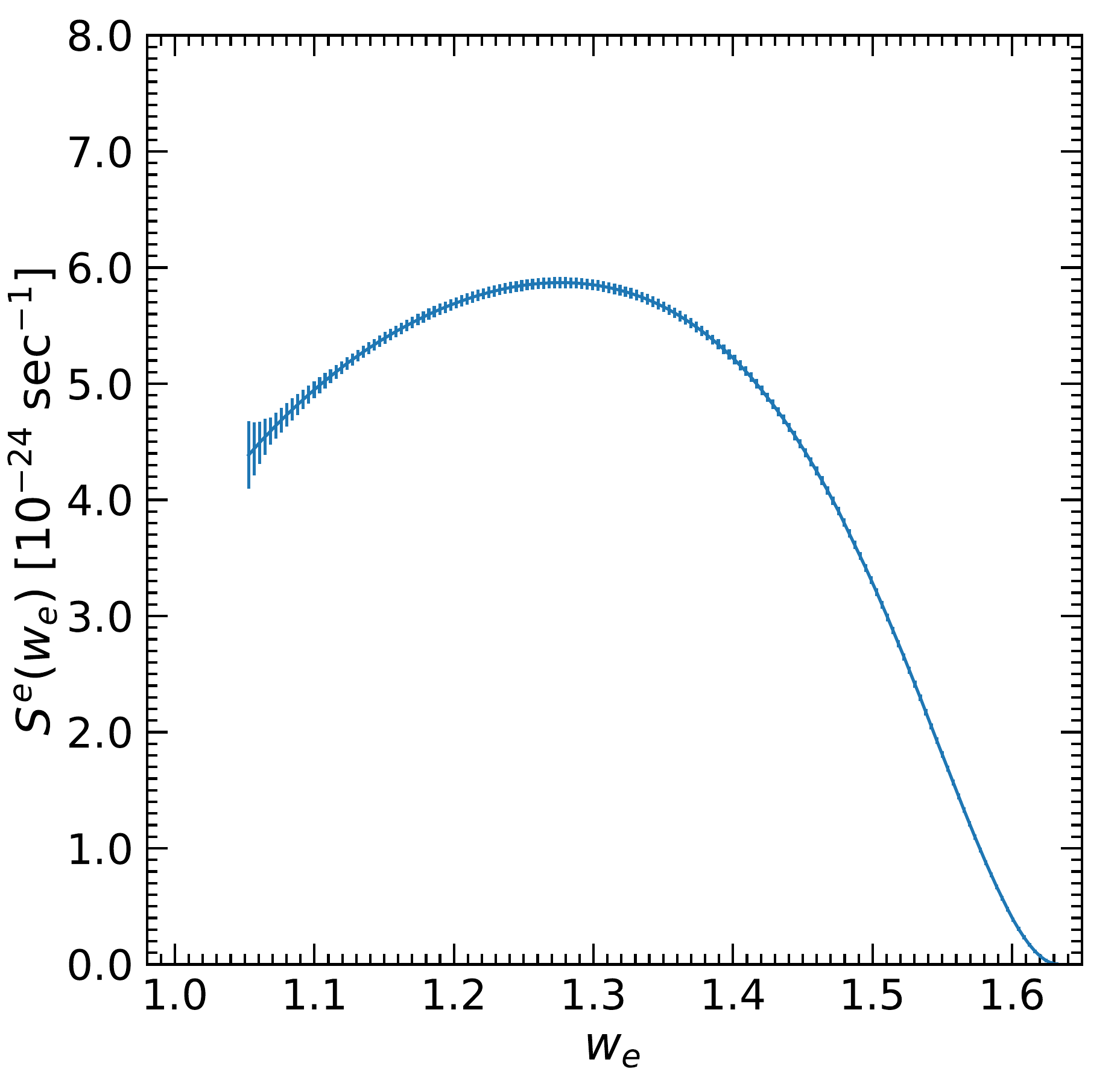}
\vspace*{-3mm}
\caption{\label{Fig_01}
\footnotesize Experimental energy spectrum for $^{113}$Cd $\beta$ decay 
(central values and total $1\sigma$ errors) as taken from \cite{Belli:2019bqp}. See the text for details.
} \end{minipage}
\end{figure}

Figure~\ref{Fig_01} shows the experimental spectrum in the range of Eq.~(\ref{exptrange}),
as taken from \cite{Belli:2019bqp} with the above normalization. The spectrum is reported in bins of width $\Delta w_e = 2~\mathrm{keV}/m_e$, together with total (statistical and systematic) uncertainty in each bin. 

From the above $S^e(w_e)$ we derive the following values for the experimental moments (up to $N=6$): 
\begin{eqnarray}
\mu^e_1 & = & 1.291\ ,\\
\mu^e_2 & = & 1.686\ ,\\
\mu^e_3 & = & 2.226\ ,\\
\mu^e_4 & = & 2.970\ ,\\
\mu^e_5 & = & 4.004\ ,\\
\mu^e_6 & = & 5.452\ .
\end{eqnarray}
We postpone the discussion of related uncertainties to Sec.~\ref{Sec:Compare}.

A final comment is in order. In \cite{Belli:2007zza}, by extrapolating the observed spectrum  below  threshold
(i.e., for $w_e\in [1,\,1.051]$), 
the total number of decays  was estimated to be $N'_e=2.40\times 10^6$ for the exposure $N_d\, t$, 
corresponding to a decay rate $\lambda = N'_e/(N_d \, t)=2.73\times 10^{-24}$~s$^{-1}$ and to
the quoted half-life $t_{1/2}=\ln 2/\lambda=8.04 \times 10^{15}$~y. As previously mentioned, 
we do not use extrapolated quantities such as $\lambda$ or $t_{1/2}$ in this work.

\subsection{Theoretical spectrum $S^t(w_e)$ and its free parameters $(r,\,s)$} 
\label{theospectra}

As discussed in \cite{Haaranen:2017ovc} and references therein, the theoretical $\beta$-decay spectrum $S^t(w_e)$ can be generally expressed as
\begin{equation}
\label{thespectrum}
S^t(w_e) = \frac{\ln2}{\kappa}\, C(w_e)\,p\,w_e\,(w_0-w_e)^2\,F_0(Z,\,w_e)\ ,
\end{equation}
where $C(w_e)$ is the so-called shape factor, $p$ is the electron momentum in units of $m_e$, and $F_0(Z,\,w_e)$ is the Fermi function with the final nuclear state having $Z=49$. In contexts where extreme precision is required, small correction terms accounting for, e.g., radiative effects and atomic screening become important. For the purposes of this work these $\approx 1\%$ corrections are insignificant but have been included as described in \cite{Haaranen:2017ovc}. The conversion constant $\kappa$ reads:
\begin{equation}
\label{kappa}
\kappa = \frac{2\pi^3 \ln 2}{m_e^5\,(G_F\,\cos\theta_C)^2}= 6289\ \mathrm{s}\ ,
\end{equation}
where $G_F$ is the Fermi constant  and  $\theta_C$ is the Cabibbo angle. For non-unique decays the shape factor has a complicated expression including universal kinematic factors and nuclear matrix elements (NME). The latter capture all the nuclear-structure dependent information regarding the decay. In the formalism of Behrens and B\"uhring \cite{Behrens1982} the NME arise from a multipole expansion of the nuclear current. The NME are then expanded as a power series resulting in an expression including vector NME ${}^V\!{\cal{M}}^{(m)}_{KLs}$ and axial-vector NME ${}^A\!{\cal{M}}^{(m)}_{KLs}$. The matrix elements with the smallest angular momenta $K$ and $L$ allowing for the decay dominate, with the first term in the power series, $m=0$, being by far the most important. For fourth-forbidden unique decays there are four leading-order NME, with the dominant matrix elements being ${}^V\!{\cal{M}}^{(0)}_{440}$ and ${}^A{\cal{M}}^{(0)}_{441}$, and with significantly smaller contributions coming from ${}^V\!{\cal{M}}^{(0)}_{431}$ and ${}^A\!{\cal{M}}^{(0)}_{541}$. The expansion can be further carried out to next-to-leading order, resulting in a total of 
45 NME \cite{Haaranen:2017ovc} that depend on $w_e$ and on powers of the nuclear radius 
$R = 1.2\, A^{1/3}\ \mathrm{fm} = 5.8\ \mathrm{fm}$.
The NME need to be numerically computed with specific nuclear models. Following 
\cite{Haaranen:2017ovc,Kostensalo:2020gha} we consider: the microscopic interacting boson-fermion model (IBFM-2), the microscopic quasiparticle-phonon model (MQPM), and the interacting shell model (ISM). 
First we discuss general aspects of the spectrum structure, and then report model-dependent numerical results in
terms of spectral moments.

In general, $C(w_e)$ is a sum over squares and products of linear combinations of NME, each being multiplied by either the vector coupling $g_{\rm V}$ or the axial-vector coupling $g_{\rm A}$. The couplings arise in the formalism of beta decays as a means to normalize the hadron current when moving from the quark level to the level of nucleons, and each axial-vector matrix element is always preceded by  $g_{\rm A}$ and each vector matrix element by $g_{\rm V}$. By defining the ratio
\begin{equation}
\label{ratio}
r = g_{\rm A}/g_{\rm V}
\end{equation}
one can formally write $S^t(w_e)$ as a quadratic form in $r$ \cite{Haaranen:2016rzs},
\begin{equation}
\label{VA}
S^t = g_{\rm V}^2(S^t_{\rm V}+r S^t_{\rm VA}+r^2 S^t_{\rm A})\ ,
\end{equation}
where $S^t_{\rm A}$ includes only axial-vector NME, $S^t_{\rm V}$ only vector NME, and $S^t_{\rm VA}$ is a mix of vector and axial-vector NME. Hereafter we shall assume as in \cite{Kostensalo:2020gha}, in accordance with the conserved vector current (CVC), that
\begin{equation}
\label{gV}
g_{\rm V} =1\ ,
\end{equation}
while $r$ will be treated as a free theoretical parameter to be fixed by the data. We shall comment on deviations from
Eq.~(\ref{gV}) in Sec.~\ref{Sec:Compare}.

As discussed in detail in \cite{Haaranen:2017ovc}, the quadratic form in Eq.~(\ref{VA}) entails delicate cancellations among large NME for 
$r\sim 1$, where agreement between theory and data can be usually found in terms of either the spectrum normalization 
\cite{Haaranen:2016rzs}
or its
shape \cite{COBRA:2018blx}, but not both at the same time (as far as $r$ is the only free parameter) \cite{Haaranen:2016rzs,Haaranen:2017ovc}. 
In particular, the main NME cancellation term turns out to be the square of a binomial, up to subleading NME terms $\epsilon$ and $\epsilon'$:
\begin{equation}
\label{cancel}
S^t \propto \left({}^V\!{\cal{M}}^{(0)}_{440} - \alpha\, {}^A\!{\cal{M}}^{(0)}_{441}\,r +\epsilon\right)^2 +\epsilon' 
\end{equation}
where $\alpha$ is a numerical coefficient of $O(1)$ and the $\cal{M}$'s are ``large'' NME, tipically of
$O(10^2)$--$(10^3)$ 
\textcolor{black}{in units of fm$^4$}; 
see 
\cite{Haaranen:2017ovc} for details of the notation and for numerical $\cal{M}$ values in the various
nuclear models. For $r\sim 1$, it turns out that the two large $\cal{M}$'s tend to cancel each other, leaving a residual of $O(1)$ \cite{Haaranen:2017ovc}.   Therefore, a subleading term $\epsilon \sim O(1)$ may still play a significant role, especially if 
its numerical value is rather uncertain. 

It was realized in \cite{Kirsebom:2019tjd,Kumar:2020vxr,Kostensalo:2020gha}
 that this role can be played by
so-called small relativistic NME, dubbed as $s$-NME in \cite{Kostensalo:2020gha} and here just as $s$ for simplicity, where:
\begin{equation}
\label{sNME}
s = {}^V\!{\cal{M}}^{(0)}_{431}\ ,
\end{equation}
%
\textcolor{black}{in units of fm$^3$}.
On the one hand, with very simple (though unrealistic) assumptions related to the nuclear-structure aspects of the decay, the CVC hypothesis would imply 
\textcolor{black}{(in our adopted units) the numerical relation 
${}^V\!{\cal{M}}^{(0)}_{431} = 0.0678 (R^{-1}) {}^V\!{\cal{M}}^{(0)}_{440}$}
\cite{Kostensalo:2020gha}, leading to 
expected values $s\sim O(1$--$10)$. On the other hand, numerical evaluations of $s$ either give $s=0$ due to model-specific limitations relating to a restricted  model space
(in the IBFM-2 and ISM models) or to $s\simeq 0.4$ (in the MQPM model) \cite{Haaranen:2017ovc}. 
\textcolor{black}{A more detailed discussion of the uncertain estimates of $s={}^V\!{\cal{M}}^{(0)}_{431}$ as compared with
${}^V\!{\cal{M}}^{(0)}_{440}$ and ${}^A\!{\cal{M}}^{(0)}_{441}$ is presented in 
Appendix~\ref{appA}.
}

Given such uncertainties, in \cite{Kostensalo:2020gha} the $s$-NME  
was simply assumed as a free parameter,  presumably of $O(1)$, to be constrained by
comparison with the data (together with $g_{\rm A}$). It turns out that, in this way, both the experimental
spectrum shape and its normalization can be well reproduced theoretically \cite{Kostensalo:2020gha}.
In the same spirit, we treat $(r,\,s)$ as free parameters in our analysis.

\begin{figure}[t!]
\begin{minipage}[c]{0.8\textwidth}
\includegraphics[width=0.8\textwidth]{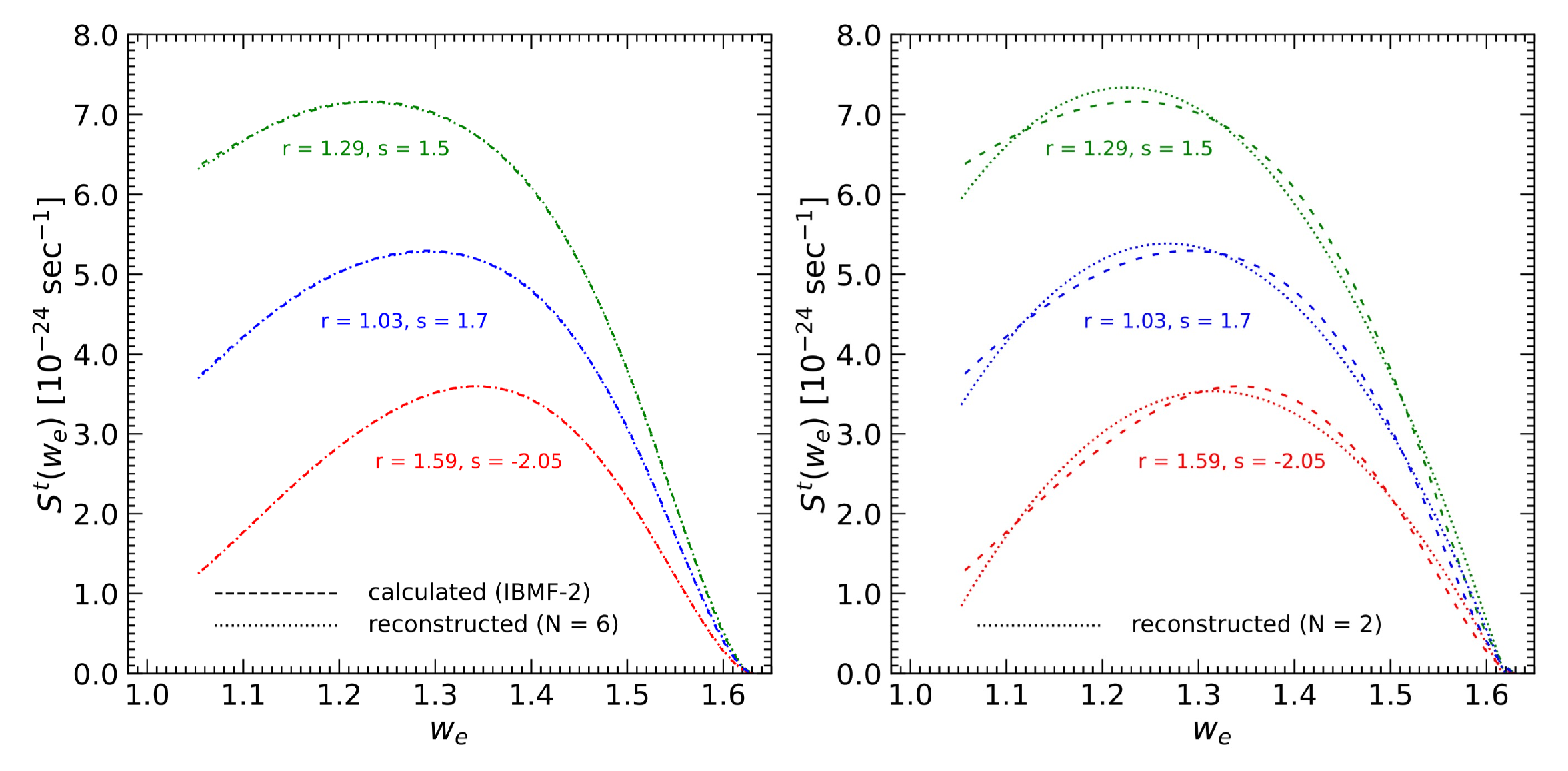}
\vspace*{-1mm}
\caption{\label{Fig_02}
\footnotesize Representative examples of theoretical spectra $S^t$ as calculated in the IBFM-2 model (dashed lines), 
for three representative pairs
of the $(r,\,s)$ parameters. The reconstruction of the spectra based on a truncated set of moments  is also shown (dotted lines)
for $N=6$ (left panel) and $N=2$ (right panel). 
} \end{minipage}
\end{figure}

Figure~\ref{Fig_02} shows three representative theoretical 
spectra $S^t(w_e)$  calculated in the IBFM-2 model (dashed lines) for three different $(r,\,s)$ values. 
Their accurate reconstruction through a set of moments truncated at $N=6$ is also shown in the left panel (dotted lines). 
Analogously, the right panel shows the reconstruction truncated at $N=2$: it can be seen that
the main qualitative features of the spectra are already captured by using just
the first three moments $\mu_0$, $\mu_1$ and $\mu_2$. Similar results hold for the spectra calculated in the MQPM and ISM models (not shown). The analysis in Sec.~\ref{Sec:Compare} 
will confirm that, in general, a few moments are enough to derive useful indications on the $(r,\,s)$ parameters.

In particular, by using as free parameters both $r$ and $s$, it appears from Fig.~\ref{Fig_02} that one can alter both the peak position and the normalization of the theoretical spectrum, and may thus hope to match the experimental spectrum in Fig.~\ref{Fig_01}. It turns out that this result is achieved for typical values $r\sim 1$ (confirming $g_{\rm A}$ quenching) and $s\sim O(1)$ (in the expected numerical ballpark), where the large NME cancellations mentioned in the context of Eqs.~(\ref{cancel},\ref{sNME}) take place.   
We shall recover very similar results, not only by using an independent data set \cite{Belli:2007zza,Belli:2019bqp} (as reported in Fig.~\ref{Fig_01}), but by adopting 
a different perspective, based on the following generalization of the quadratic form in Eq.~(\ref{VA}).

We observe that, since $s$ adds to the NME terms multiplied by $g_{\rm V}$, it must appear up to second power
in $S^t(\omega)$. Moreover, a mixed dependence $\propto r\cdot s$ 
must emerge from the VA term in Eq.~(\ref{VA}).  
\textcolor{black}{Therefore, the spectrum $S^t(w_e)$ must be a quadratic form  in both $r$ and $s$, as well as any
integral over it involving $w_e^n$ (with $n\geq 0$),
\begin{equation}
\label{quadra}
\int_{w_{\rm thr}}^{w_0} S^t(w_e)\,w_e^n\,dw_e = \sum_{i+j\leq 2}a^n_{ij}\, r^i\, s^j \ ,
\end{equation}
where the numerical coefficients $a^n_{ij}$  (with $n$ being a superscript, not a power) are expressed in units of s$^{-1}$ as the spectrum $S^t$. 
The $0^{\rm th}$ moment $\mu^t_0(r,\,s)$ corresponds to the above quadratic form with $n=0$,
while the  $n^{\rm th}$ moment $\mu^t_n(r,\,s)$ for $n\geq 1$ corresponds to a ratio of quadratic forms (with 
index $n$ at numerator and zero at denominator).
}

In practice, to evaluate the  $a^n_{ij}$  coefficients for a given nuclear model at fixed $n$, 
one just needs to calculate the energy spectrum $S^t(w_e\,|\,r,\,s)$ at
six arbitrary points $(r,\,s)$, evaluate the integral on the l.h.s.\ of Eq.~(\ref{quadra}), and solve  
in the six unknowns $\{a^n_{ij}\}$.%
\footnote{ 
To be sure, we have numerically checked the validity of Eq.~(\protect\ref{quadra}) 
over a dense grid sampling the relevant $(r,\,s)$ parameter space, in all of the three nuclear models 
(IBFM-2, MQPM and ISM).
}
 Table~\ref{Tab1} reports the numerical $\{a^n_{ij}\}$ values (in units of $10^{-24}$~s$^{-1}$) up to $N=6$,
for the three nuclear models discussed in this work.

\begin{table}[bh!]
\centering
\resizebox{.75\textwidth}{!}{\begin{minipage}{\textwidth}
\caption{\label{Tab1} 
Coefficients $a^n_{ij}$ of the quadratic forms parametrizing the theoretical moments up to $N=6$, in each of the three 
nuclear models considered in this work. The $a^n_{ij}$ are expressed in units of $10^{-24}$~s$^{-1}$.  
}
\begin{ruledtabular}
\begin{tabular}{lccccccc}
Model & $n$ & $a^n_{20}$ & $a^n_{02}$ & $a^n_{11}$ & $a^n_{10}$  & $a^n_{01}$ & $a^n_{00}$ \\
\hline
    	& 0 & $+2.8998$ & $+0.7914$ & $+2.8163$ & $-5.8662$ & $-2.9952$ & $+3.1087$  \\		
    	& 1 & $+3.6046$ & $+1.0225$ & $+3.5722$ & $-7.2731$ & $-3.7984$ & $+3.8552$  \\
    	& 2 & $+4.5331$ & $+1.3358$ & $+4.5835$ & $-9.1214$ & $-4.8722$ & $+4.8357$  \\		
 IBFM-2 & 3 & $+5.7690$ & $+1.7640$ & $+5.9492$ & $-11.574$ & $-6.3214$ & $+6.1368$  \\
	    & 4 & $+7.4305$ & $+2.3542$ & $+7.8102$ & $-14.862$ & $-8.2949$ & $+7.8803$  \\		
    	& 5 & $+9.6855$ & $+3.1738$ & $+10.368$ & $-19.312$ & $-11.005$ & $+10.239$  \\
    	& 6 & $+12.774$ & $+4.3204$ & $+13.912$ & $-25.390$ & $-14.758$ & $+13.460$  \\
\hline
    	& 0 & $+20.086$ & $+0.7914$ & $+7.5565$ & $-41.059$ & $-7.8113$ & $+21.146$  \\
    	& 1 & $+24.911$ & $+1.0225$ & $+9.5846$ & $-50.905$ & $-9.9058$ & $+26.223$  \\
	    & 2 & $+31.253$ & $+1.3358$ & $+12.298$ & $-63.839$ & $-12.706$ & $+32.892$  \\
  MQPM  & 3 & $+39.676$ & $+1.7640$ & $+15.962$ & $-81.005$ & $-16.485$ & $+41.740$  \\		
    	& 4 & $+50.976$ & $+2.3542$ & $+20.955$ & $-104.01$ & $-21.632$ & $+53.598$  \\
    	& 5 & $+66.281$ & $+3.1738$ & $+27.817$ & $-135.15$ & $-28.700$ & $+69.640$  \\	
    	& 6 & $+87.205$ & $+4.3204$ & $+37.325$ & $-177.68$ & $-38.485$ & $+91.542$  \\
\hline
    	& 0 & $+17.509$ & $+0.7914$ & $+7.0682$ & $-33.572$ & $-6.8313$ & $+16.166$  \\
    	& 1 & $+21.713$ & $+1.0225$ & $+8.9655$ & $-41.627$ & $-8.6632$ & $+20.050$  \\
	    & 2 & $+27.238$ & $+1.3358$ & $+11.504$ & $-52.208$ & $-11.113$ & $+25.151$  \\
  ISM   & 3 & $+34.576$ & $+1.7640$ & $+14.932$ & $-66.252$ & $-14.418$ & $+31.919$  \\		
    	& 4 & $+44.418$ & $+2.3542$ & $+19.603$ & $-85.078$ & $-18.920$ & $+40.990$  \\
    	& 5 & $+57.748$ & $+3.1738$ & $+26.024$ & $-110.56$ & $-25.103$ & $+53.262$  \\	
    	& 6 & $+75.969$ & $+4.3204$ & $+34.920$ & $-145.36$ & $-33.663$ & $+70.018$  \\
\end{tabular}
\end{ruledtabular}
\end{minipage}}
\end{table}

Summarizing, we have discretized the continuous information contained in the infinite family of 
spectra $S^t(w_e\,|\,r,\,s)$ into a small number of moments $\mu_n^t$, 
each depending on simple quadratic forms in the free parameters $r$ and $s$
(involving six coefficients $a^n_{ij}$ at any $n$). This approach greatly simplifies the numerical calculation
of the theoretical moments, as well as their comparison with the experimental moments $\mu_n^e$, 
as discussed below.

\textcolor{black}{\section{Spectral moment method: Comparison of theory and data} \label{Sec:Compare}} 

In this Section we explore the implications of equating a finite set of theoretical and experimental moments:
\begin{equation}
\label{impose}
\mu^t_n(r,\,s) = \mu^e_n \ (n=0,\,1,\dots N)\ .
\end{equation}

Each of the above equations sets a quadratic form in $(r,\,s)$ equal to a constant, and thus leads to a conic section in the corresponding coordinates. It turns out that, for $n=0$, the conic section is a slanted and elongated ellipse, while for 
$n\geq 1$ the conics form a bundle of hyperbolae. In the ideal case (perfect match between theory and data), all these curves would intersect in a single $(r,\,s)$ point; in real cases, the various crossing points will cluster with some dispersion around a preferred $(r,\,s)$ region.  The smaller the dispersion, the better the agreement between the experimental and theoretical moments and spectra. In Appendix~\ref{appB} we discuss general features of the conic sections and of their crossings, that allow to visualize the effects of large NME cancellations, as well as to interpret previous fit results obtained in \cite{Kostensalo:2020gha} through the 
revised spectrum shape method.
Below we show our results in the $(r,\,s)$ plane and discuss the preferred parameter values in the three nuclear models considered.

\begin{figure}[b!]
\begin{minipage}[c]{0.99\textwidth}
\includegraphics[width=0.99\textwidth]{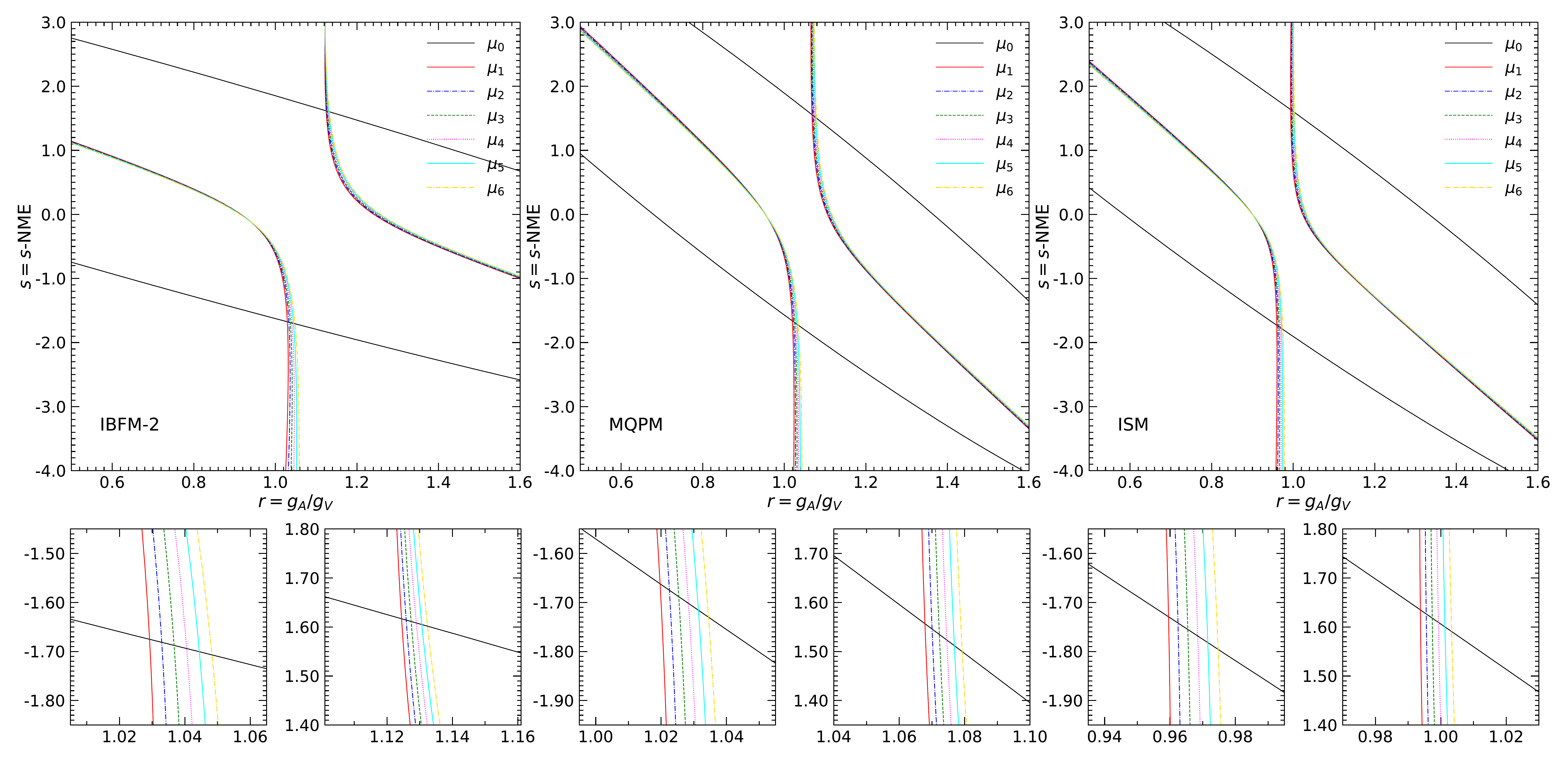}
\vspace*{-2mm}
\caption{\label{Fig_03}
\footnotesize  Numerical results obtained by imposing the equality of theoretical and experimental moments, $\mu_n^t=\mu_n^e$ (up to $N=6$) in the plane charted by the
free parameters $r=g_{\rm A}/g_{\rm V}$ and $s=s$-NME.  The left, middle and right plots correspond to the IBFM-2, MQPM and ISM models, respectively. In each of the three upper panels, note (part of) the slanted ellipse determined by the zeroth moment, and the bundle of hyperbolae determined by the first and higher moments. For each model, the two lower panels zoom in the regions at $s<0$ (left) and $s>0$ (right) where the ellipse and the bundle cross each other. See the text for details.   
}
\end{minipage}
\end{figure}

Figure~\ref{Fig_03} shows the {loci of points in the $(r,\,s)$ plane fulfilling Eq.~(\ref{impose})} up to $N=6$,
 for the  model IBMF-2 (left), MQPM (middle), and ISM (right). In each panel, one can see (part of) the slanted ellipse determined by the zeroth moment, and the bundle of hyperbolae determined by the first and higher moments. 
The two regions where the ellipse and the bundle cross each other correspond to positive and negative values of $s$, and are enlarged in the lower set of panels. As discussed in Appendix~\ref{appB}, in principle there are
two other regions of crossing, close to the extremal sides of the ellipse and thus beyond scale (not shown), that would correspond to unphysical values of $r$ (much smaller or much larger than unity).

In each of the enlarged crossing regions reported in Fig.~\ref{Fig_03} (lower panels), 
the bundle of hyperbolae shows some dispersion, that turns out to be smaller for $s>0$ as compared with $s<0$, and minimal for the IBFM-2 model. Therefore, we expect that the experimental spectrum is best matched by the theoretical spectra at $s>0$ (w.r.t.~$s<0$), in particular for the
IBFM-2 model. For definiteness, we check these expectations by calculating 
the $S^t$ spectra at the points where the $\mu_0$ ellipse intersects the  $\mu_1$ hyperbola, whose coordinates  
as reported in Table~\ref{Tab2}.

\begin{table}[t!]
\centering
\resizebox{.5\textwidth}{!}{\begin{minipage}{.5\textwidth}
\caption{\label{Tab2} \footnotesize 
Intersection points of the 0$^{\rm th}$ and 1$^{\rm st}$ moment curves in the $(r,\,s)$ plane, for each of the three nuclear models considered.  
}
\begin{ruledtabular}
\begin{tabular}{lcc}
Model & $r$ & $s$  \\
\hline
IBFM-2    	& 1.125 & $+1.617$    \\		
    		& 1.029 & $-1.675$    \\		
\hline		
MQPM    	& 1.068 & $+1.557$    \\		
    		& 1.020 & $-1.662$    \\		
\hline
ISM    		& 0.994 & $+1.635$    \\		
    		& 0.960 & $-1.729$    \\
\end{tabular}
\end{ruledtabular}
\end{minipage}}
\end{table}

\begin{figure}[b!]
\begin{minipage}[c]{0.99\textwidth}
\includegraphics[width=0.98\textwidth]{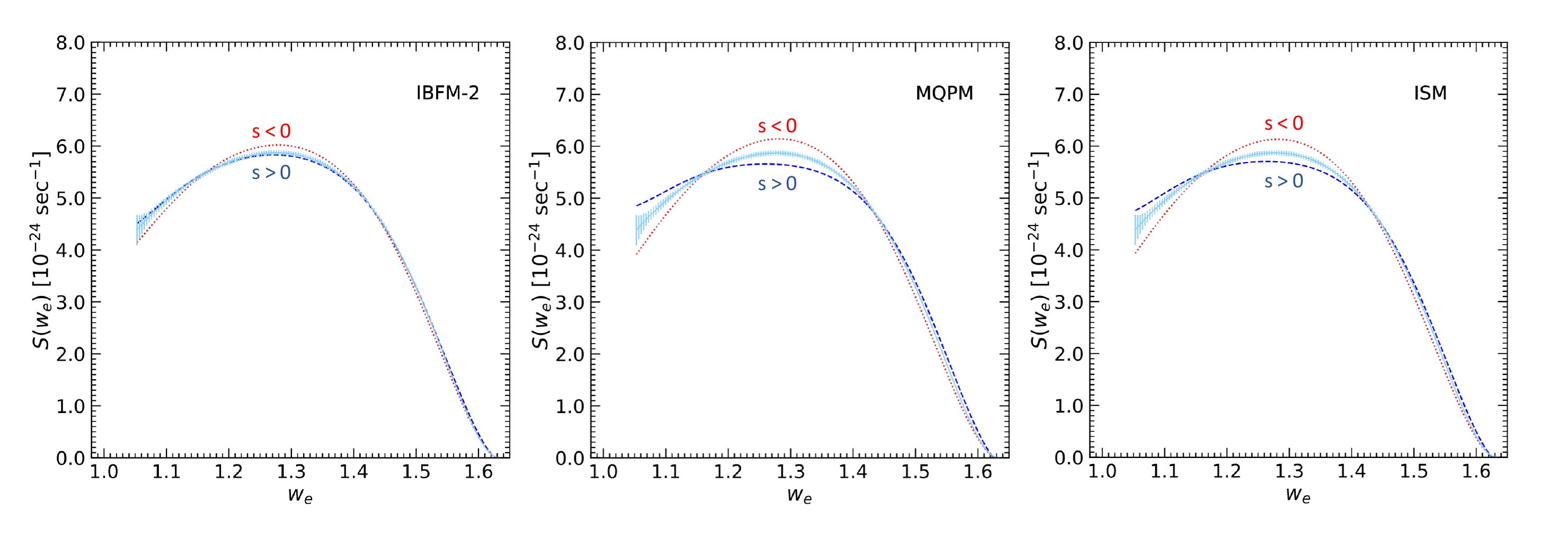}
\vspace*{-3mm}
\caption{\label{Fig_04}
\footnotesize  Theoretical spectra calculated at the $(r,\,s)$ points reported in Table~\ref{Tab2}, for each of the three nuclear models IBMF-2, MQPM and ISM (left, middle and right panels), as compared with the experimental spectrum of Fig.~\ref{Fig_01}. 
In each panel, the dashed blue curve refers to the case $s>0$, and the dotted red curve to the case $s<0$.
} 
\end{minipage}
\end{figure}

Figure~\ref{Fig_04} shows the theoretical spectra corresponding to the above $(r,\,s)$ values 
for the three nuclear models: IBFM-2 (left), MQPM (middle) and ISM (right). In each panel, the experimental spectrum (in light blue)
should be compared with the blue-dashed and red-dotted spectra, referring 
to positive and negative values of $s$ in Table~\ref{Tab2}, respectively. The spectra with $s>0$ are generally slightly broader and less peaked than the experimental spectrum, while the opposite happens for the spectra with $s<0$. The deviations from the experimental spectrum are generally smaller for $s>0$ than for $s<0$. and can be as small as the experimental errors for the IBFM-2 model at $s>0$. 
In this context, we recall that the model IBFM-2 predicts a priori $s=0$ \cite{Haaranen:2017ovc}, and that only
by fixing $s\sim 1.6$ one gets the good agreement with the experimental data in Fig.~\ref{Fig_04}; 
see Appendix~\ref{appA} for details.

We thus find that all models point towards $r\sim 1$, corresponding to a
quenching factor $q=g_{\rm A}/g_{\rm A}^\mathrm{vac}\sim 0.8$ assuming $g_{\rm V}=1$; and towards $|s|\simeq 1.6$, corresponding to a small 
vector-like relativistic NME in the expected ballpark of $O(1)$, with a preference for $s>0$. More precisely, by grouping the values
in Table~\ref{Tab2},
we find the best match between theory and data around
\begin{eqnarray}
r &\simeq 0.99  -  1.13\ ,\label{summary1} \label{sum1}\\ 
s &\simeq 1.56  -  1.64\ ,\label{summary2} \label{sum2}
\end{eqnarray}
with a secondary (worse) match around $r\simeq 0.96$--$1.03$ and $s=-(1.66$--$1.73)$, that cannot be excluded a priori from a  phenomenological viewpoint (see also Appendix~\ref{appA}). Remarkably, the above ranges correspond to 
relatively small uncertainties on the $(r,\,s)$ parameters.

It would be tempting to refine the indications in favor of $r\sim 1$ and $|s|\sim 1.6$ by 
attaching more accurate error estimates to these parameters, as derived by 
detailed data fits including both experimental and theoretical uncertainties. 
However, in our case 
the theoretical shape errors are likely to be larger than the data errors  (in contrast with the data analysis
in \cite{Kostensalo:2020gha}), as suggested by the fact that, in Fig.~\ref{Fig_04}, the theoretical spectra
are generally outside (or at border of) the experimental error band. As a check, we have performed  
numerical least-squares adjustments of the theoretical spectra 
by including only experimental uncertainties, either by fitting the binned spectrum
with uncorrelated total errors in Fig.~\ref{Fig_01} or by fitting the associated $N=6$ moments with their propagated covariance matrix.
In both cases we obtain 
unreasonably high values of the minimum $\chi^2$ per d.o.f. (except for the noted IBFM-2 case at $s>0$), 
and unreasonably tiny errors on $(r,\,s)$ at subpercent level (in all cases), that are much smaller
than the realistic few-percent spread of the same parameters  (see Table~\ref{Tab2}).
On the positive side, by using only experimental errors, 
the $(r,\,s)$ best fits  are invariably close to the $s>0$ solutions reported in Eqs.~(\ref{sum1},\ref{sum2})
and their associated spectra are close to those in Fig.~\ref{Fig_04}, within percent-level deviations (not shown). 
The SMM appears thus to provide reasonably correct and robust $(r,\,s)$ values,
even with a limited amount of information and without the need for detailed data fits.
Future improvements may include theoretical spectrum 
shape errors as estimated, e.g., by a detailed analysis of the inputs or approximations inherent the 
current next-to-leading order NME calculations in 
each nuclear model: a task that is beyond the scope 
of this work.

Similar results on $(r,\,s)$ were obtained in  \cite{Kostensalo:2020gha}, 
by applying a revised spectrum shape approach to an independent set of $^{113}$Cd $\beta$-decay data, characterized
by a higher threshold ($T_{\rm thr}\simeq 52$--132~keV in different detectors, with $\langle T_{\rm thr}\rangle = 92$~keV) with respect to the data used herein \cite{Belli:2007zza} (having a single $T_{\rm thr}=26$~keV). In \cite{Kostensalo:2020gha}, the
preferred values for the free parameters were found to cluster around $g_{\rm A}\simeq 0.83$--$0.99$ 
(somewhat lower than our $g_{\rm A}\sim 0.99$--1.13) and 
around $s\simeq 1.85$--2.1 (somewhat higher than our $s\simeq 1.56$--1.64). We surmise that these differences may be due in part to
the different $^{113}$Cd datasets  and in part to the different approach used in the analysis.  
In particular, in this work the spectrum normalization is constrained through the 0$^{\rm th}$ moment (i.e., the absolute decay rate $\mu_0$ above the 16~keV threshold), while
 in  \cite{Kostensalo:2020gha} it is constrained through the decay half life (that requires a spectrum extrapolation 
below each of the 52--132~keV detector thresholds). As noted at the end of Section~\ref{notation}, such
extrapolations may lead to biases. Altering the normalization leads to anticorrelated changes 
between $g_{\rm A}$ and $s$ (as suggested by the $\mu_0$ ellipse in our approach), which is what we find in comparison 
with \cite{Kostensalo:2020gha}. Apart from these differences, a robust message
emerges from \cite{Kostensalo:2020gha} and from this work: quenched values of the axial-vector coupling 
(around $g_{\rm A}\simeq 0.9$--1), 
accompanied by an adjustment of the small vector NME  (around $s\simeq 1.6$--2), 
are required to match the existing $^{113}$Cd $\beta$-decay
spectrum data in both normalization and shape, in each of the three nuclear models considered.

Around these $(r,\,s)$ values, large NME cancellations take place,
leaving residual spectra that reasonably reproduce the experimental spectrum both in normalization (0$^{\rm th}$ moment) and in
shape (1$^{\rm st}$ and higher moments). These phenomenological facts strongly suggest that all the models
require the following adjustments: 
an overall multiplicative correction to the axial-vector NME, via a quenched value $g_{\rm A}\sim 1$ (for $g_V=1$);  and 
a small additive vector-like correction, parametrized by a $s$-NME with an absolute value around $s\sim 1.6$.
One might wonder whether $g_{\rm V}$ could be used as free parameters instead of $s$, abandoning the CVC assumption 
$g_V=1$. 
The answer is negative, since the prefactor $g^2_{\rm V}$ in Eq.~(\ref{VA}) would only affect the zeroth moment $\mu_0$ 
but would cancel out in the first and higher moments [Eq.~(\ref{moment})], leading to a spectrum shape depending 
on a single free parameter $r$. It was already concluded in \cite{Haaranen:2016rzs} 
that one cannot match both the spectrum normalization and its shape by varying just $g_{\rm V}$ besides $g_{\rm A}$. In other words,  
the additive adjustment parametrized by $s$ cannot be traded for a multiplicative adjustment parametrized by $g_{\rm V}$. 
At present, the CVC assumption $g_{\rm V}=1$ can thus be safely maintained in the context of $^{113}$Cd $\beta$ decay.

\vspace*{9mm}
\section{Summary and perspectives \label{Sec:Summary}}

In this work we have studied the normalization and shape of the electron
energy spectrum of the fourth-forbidden $\beta$ decay of $^{113}$Cd with a novel approach, coined the spectral-moment method (SMM), based on a truncated set of
spectral moments $\mu_n$.
The 0$^{\rm th}$ moment is related to the normalization, while the 1$^{\rm st}$ and higher moments are related to the shape. 
As in \cite{Kostensalo:2020gha}, we have assumed that the spectra depend on two free parameters: an axial-vector coupling parameter 
$r=g_{\rm A}/g_{\rm V}$ (for $g_{\rm V}=1)$ and a small relativistic vector-like NME parameter $s$. We have shown that: each moment is a quadratic 
form in $(r,\,s)$; iso-moment curves are ellipses for $\mu_0$ and hyperbolae for $\mu_{1,2,3\dots}$; the intersections of
a few moment curves are enough to derive interesting constraints on $(r,\,s)$, without detailed data fits 
\textcolor{black}{%
(see also Appendix B).}

In particular, by equating the theoretical moments with the experimental ones, as derived from the data in \cite{Belli:2019bqp},
the following results emerge: the intersection of the $\mu_0$ and $\mu_1$ curves provides $r\sim 1$ and $|s|\sim 1.6$; the 
case $s>0$ results in a smaller spread of intersections with higher-moment curves and is thus preferred, as also confirmed
by visual inspection of the spectra.  
\textcolor{black}{%
Nuclear model considerations also suggest $s\geq 0$, although  $s<0$ cannot be excluded a priori
(see also Appendix~\ref{appA}).} 
The spread of the $(r,\,s)$ values in Table~\ref{Tab2}, at the level 
of a few percent at least, exceeds the purely experimental uncertainties and call for (currently unquantified) theoretical 
spectrum-shape uncertainties. In any case, our results are in the same ballpark of those obtained in
\cite{Kostensalo:2020gha} with a different methodology and independent data. In general, 
the derived $(r,\,s)$ values provide evidence for a multiplicative renormalization (quenching) of the axial coupling $g_{\rm A}$ and for 
an additive adjustment of vector-like terms via the small relativistic $s$-NME.

\newpage
Since the main quantitative information (apart from indications about the sign of $s$) has been derived just from $\mu_0$ and $\mu_1$,
we surmise that the SMM can be quite powerful even when the experimental data are less accurate than those used in this work for $^{113}$Cd. In this context, it should be noted that $^{113}$Cd is just one of several nuclei where 
forbidden $\beta$ decays occur with a significant spectral dependence on $g_{\rm A}$ \cite{Haaranen:2016rzs,Kostensalo:2017jgw,Kostensalo:2017xxq,Kumar:2021euw} 
and possibly on other parameters such as the $s$-NME or similar ones. 

In cases where the available spectral data are scarce, it should anyway be possible to derive, within a specified energy
window $w_e\in [w_e^{\min},\,w_e^{\max}]$, at least the spectrum normalization $(\mu_0)$ and the average energy $(\mu_1)$
with reasonable approximation.  By equating the experimental and theoretical values for these two moments, constraints on 
$(r,\,s)$ or equivalent parameters could be derived, without the need of complicated data fits. 
Among the forbidden $\beta$-decay spectra with significant $g_{\rm A}$ dependence, of particular importance is the
$^{115}$In spectrum \cite{Haaranen:2016rzs}, that was experimentally observed long ago in \cite{OldLi} and 
recently measured with a bolometric detector in \cite{Leder:2022beq}. A first data analysis with the spectrum-shape method (SSM),
and using only $g_{\rm A}$ as a free parameter, suggests significant $g_{\rm A}$ quenching \cite{Leder:2022beq} but does not account 
for both normalization and shape at the same time. It remains to be seen if allowance for an extra parameter such as the $s$-NME
can provide a match to all data. The presently introduced SMM might allow a rapid check of this possibility, and will be
applied to, e.g., existing $^{115}$In data in a separate work.

\vspace*{-3mm}
\acknowledgments

The work of E.L.\ and A.M.\ is partly supported the Italian Ministero dell'Universit\`a e Ricerca (MUR) through the research grant number 2017W4HA7S ``NAT-NET: Neutrino and Astroparticle Theory Network'' under the program PRIN 2017, and by the Istituto Nazionale di Fisica Nucleare (INFN) through the``Theoretical Astroparticle Physics'' (TAsP) project. 
We are grateful to P.~Belli, R.~Bernabei, F.~A.~Danevich and V.~I.~Tretyak for providing us with a digitized form of 
the $^{113}$Cd $\beta$-decay spectrum data \cite{Belli:2007zza} as updated in Fig.~29 of \cite{Belli:2019bqp}, and for useful correspondence.

\vspace*{-4mm}
\appendix
\textcolor{black}{\section{Analysis of relevant NME} \label{appA}}

In the Tables~\ref{tab:A1}--\ref{tab:A3} we have listed the single-particle (s.p.) matrix elements corresponding to all relevant transitions in the vector and axial-vector NME of interest for our analysis, namely, the small $s={}^V\!{\cal{M}}^{(0)}_{431}$ and the large (and largely cancelling) terms  ${}^V\!{\cal{M}}^{(0)}_{440}$ and ${}^A\!{\cal{M}}^{(0)}_{441}$. The relevant pre-factors have been included so that these correspond to single-particle model NME and are thus comparable to the numerical NME values listed in \cite{Haaranen:2017ovc} for $^{113}$Cd.

\begin{table}[b]
\centering
\resizebox{.45\textwidth}{!}{\begin{minipage}{0.56\textwidth}
\caption{\label{tab:A1} 
The single-particle matrix elements ${}^V\!{\cal{M}}^{(0)}_{431}$ [fm$^3$] for $^{113}$Cd for the relevant orbitals.}
    \begin{ruledtabular}
    \begin{tabular}{rrrrrrrrrrr}
 & 	$0f_{7/2}$	&	$0f_{5/2}$	&	$1p_{3/2}$	&	$1p_{1/2}$	&	$0g_{9/2}$	&	$0g_{7/2}$	&	$1d_{5/2}$	&	$1d_{3/2}$	&	$2s_{1/2}$	&	$0h_{11/2}$	\\
\hline
$0f_{7/2}$&	0	&	0	&	0	&	0	&	0	&	0	&	0	&	0	&	0	&	9.5	\\
$0f_{5/2}$&	0	&	0	&	0	&	0	&	0	&	0	&	0	&	0	&	0	&	-4.3	\\
$1p_{3/2}$&	0	&	0	&	0	&	0	&	0	&	0	&	0	&	0	&	0	&	-12.5	\\
$1p_{1/2}$&	0	&	0	&	0	&	0	&	0	&	0	&	0	&	0	&	0	&	0	\\
$0g_{9/2}$&	0	&	0	&	0	&	0	&	0	&	0	&	0	&	0	&	0	&	0	\\
$0g_{7/2}$&	0	&	0	&	0	&	0	&	0	&	0	&	0	&	0	&	0	&	0	\\
$1d_{5/2}$&	0	&	0	&	0	&	0	&	0	&	0	&	0	&	0	&	0	&	0	\\
$1d_{3/2}$&	0	&	0	&	0	&	0	&	0	&	0	&	0	&	0	&	0	&	0	\\
$2s_{1/2}$&	0	&	0	&	0	&	0	&	0	&	0	&	0	&	0	&	0	&	0	\\
$0h_{11/2}$&	-9.5	&	-4.3	&	12.5	&	0	&	0	&	0	&	0	&	0	&	0	&	0	
\end{tabular}
\end{ruledtabular}
\end{minipage}}
\end{table}

Given that the initial and final states are ground states with spin-parities $1/2^+$ and $9/2^+$, respectively, the transition is most likely dominated by a transition between the neutron orbital $2s_{1/2}$ and the proton orbital $0g_{9/2}$. This means ${}^V\!{\cal{M}}^{(0)}_{431}$(s.p.)$=0$ fm$^3$, ${}^V\!{\cal{M}}^{(0)}_{440}$(s.p.)$=637$ fm$^4$, and ${}^A\!{\cal{M}}^{(0)}_{441}$(s.p.)$=565$ fm$^4$. The nuclear models give  ${}^V\!{\cal{M}}^{(0)}_{440}=$317--827 fm$^4$, and ${}^A\!{\cal{M}}^{(0)}_{441}=$314--848 fm$^4$, which are in reasonable agreement with the single-particle NME with the final values depending on contributions from other near-by orbitals. For ${}^V\!{\cal{M}}^{(0)}_{431}$ the situation is however very different, as no orbitals near the Fermi-surface have nonzero contributions. Furthermore, the CVC relation ${}^V\!{\cal{M}}^{(0)}_{431} = 0.0678 (R^{-1}) {}^V\!{\cal{M}}^{(0)}_{440}$ mentioned in Sec.~\ref{theospectra}
(which does not apply exactly when multiple configurations are allowed for the wave functions) suggests that ${}^V\!{\cal{M}}^{(0)}_{431}$(MQPM, CVC) = 9.7 fm$^3$, ${}^V\!{\cal{M}}^{(0)}_{431}$(IBFM-2, CVC)=3.7 fm$^3$, and ${}^V\!{\cal{M}}^{(0)}_{431}$(ISM) = 8.4 fm$^3$, while the models give the values 0.37 fm$^3$, 0 fm$^3$, and 0 fm$^3$, respectively. For the CVC values for the smallest NME to hold, the transition would need to be basically a pure $0h_{11/2}$-$1p_{3/2}$ which does not seem reasonable based on the spin-parities, the facts that these are ground states, and the proton and neutron numbers of the nuclei.

\newpage
\begin{table}[t]
\centering
\resizebox{.45\textwidth}{!}{\begin{minipage}{0.56\textwidth}
\caption{\label{tab:A2} 
The single-particle matrix elements ${}^V\!{\cal{M}}^{(0)}_{440}$ [fm$^4$] for $^{113}$Cd for the relevant orbitals.}
    \centering
    \begin{ruledtabular}
    \begin{tabular}{rrrrrrrrrrr}
&	$0f_{7/2}$	&	$0f_{5/2}$	&	$1p_{3/2}$	&	$1p_{1/2}$	&	$0g_{9/2}$	&	$0g_{7/2}$	&	$1d_{5/2}$	&	$1d_{3/2}$	&	$2s_{1/2}$	&	$0h_{11/2}$	\\
\hline
$0f_{7/2}$&	434	&	319	&	-363	&	-426	&	0	&	0	&	0	&	0	&	0	&	695	\\
$0f_{5/2}$&	-319	&	336	&	484	&	0	&	0	&	0	&	0	&	0	&	0	&	-306	\\
$1p_{3/2}$&	-363	&	-484	&	0	&	0	&	0	&	0	&	0	&	0	&	0	&	-907	\\
$1p_{1/2}$&	426	&	0	&	0	&	0	&	0	&	0	&	0	&	0	&	0	&	0	\\
$0g_{9/2}$&	0	&	0	&	0	&	0	&	729	&	382	&	-644	&	-397	&	637	&	0	\\
$0g_{7/2}$&	0	&	0	&	0	&	0	&	-382	&	621	&	452	&	-501	&	-570	&	0	\\
$1d_{5/2}$&	0	&	0	&	0	&	0	&	-644	&	-452	&	583	&	820	&	0	&	0	\\
$1d_{3/2}$&	0	&	0	&	0	&	0	&	397	&	-501	&	-820	&	0	&	0	&	0	\\
$2s_{1/2}$&	0	&	0	&	0	&	0	&	637	&	570	&	0	&	0	&	0	&	0	\\
$0h_{11/2}$&	695	&	306	&	-907	&	0	&	0	&	0	&	0	&	0	&	0	&	1111	
\end{tabular}
\end{ruledtabular}
\vspace*{-14mm}
\end{minipage}}
\end{table}

\phantom{.}
\begin{table}[h!]
\centering
\resizebox{.45\textwidth}{!}{\begin{minipage}{0.56\textwidth}
\caption{\label{tab:A3} 
The single-particle matrix elements ${}^A\!{\cal{M}}^{(0)}_{441}$ [fm$^4$] for $^{113}$Cd for the relevant orbitals.}
    \begin{ruledtabular}
    \begin{tabular}{rrrrrrrrrrr}
&	$0f_{7/2}$	&	$0f_{5/2}$	&	$1p_{3/2}$	&	$1p_{1/2}$	&	$0g_{9/2}$	&	$0g_{7/2}$	&	$1d_{5/2}$	&	$1d_{3/2}$	&	$2s_{1/2}$	&	$0h_{11/2}$	\\
\hline
$0f_{7/2}$&	0	&	500	&	-160	&	-475	&	0	&	0	&	0	&	0	&	0	&	-301	\\
$0f_{5/2}$&	500	&	0	&	-538	&	0	&	0	&	0	&	0	&	0	&	0	&	615	\\
$1p_{3/2}$&	160	&	-538	&	0	&	0	&	0	&	0	&	0	&	0	&	0	&	794	\\
$1p_{1/2}$&	-475	&	0	&	0	&	0	&	0	&	0	&	0	&	0	&	0	&	0	\\
$0g_{9/2}$&	0	&	0	&	0	&	0	&	0	&	774	&	-283	&	-624	&	565	&	0	\\
$0g_{7/2}$&	0	&	0	&	0	&	0	&	774	&	0	&	-704	&	222	&	632	&	0	\\
$1d_{5/2}$&	0	&	0	&	0	&	0	&	283	&	-704	&	0	&	911	&	0	&	0	\\
$1d_{3/2}$&	0	&	0	&	0	&	0	&	-624	&	-222	&	911	&	0	&	0	&	0	\\
$2s_{1/2}$&	0	&	0	&	0	&	0	&	-565	&	632	&	0	&	0	&	0	&	0	\\
$0h_{11/2}$&	301	&	615	&	-794	&	0	&	0	&	0	&	0	&	0	&	0	&	0	
\end{tabular}
\end{ruledtabular}
\end{minipage}}
\end{table}

For ISM and IBFM-2 the value of ${}^V\!{\cal{M}}^{(0)}_{431}$ is systematically zero, as in the ISM calculations the orbital $0h_{11/2}$ was kept empty to reduce the large computational burden, and in IBFM-2 the initial state with spin-parity $1/2^+$ cannot be formed by pairing $0^+$ and $2^+$ bosons with a fermion with spin-parity $11/2^-$. For MQPM the contributions between $0h_{11/2}$ and the lower proton orbitals are included, but the contributions for the higher orbitals are not very reliable as the parameter tuning can be reasonably done only for the lowest orbitals. 

Based on these arguments, we surmise that the $s$-NME is not well described by the nuclear models while the other NME do not suffer from the same problems. Therefore, it makes sense to take the $s$-NME as a tuning parameter instead of the large matrix elements. 
Concerning the sign of $s$, the option $s>0$ is theoretically regarded as being more reasonable \cite{Kostensalo:2020gha}, since all the above estimates---despite being largely uncertain---typically provide $s\geq 0$. However, one cannot exclude a priori that $s<0$,
also because some single-particle contributions may be negative (see Table~\ref{tab:A1}).

\vspace*{-5mm}
\textcolor{black}{\section{Quadratic forms in the $(r,\,s)$ parameters and conic sections} \label{appB}}

As mentioned in Sec.~\ref{theospectra} and discussed in 
\cite{Kostensalo:2020gha}, in the $(r,\,s)$ region where theoretical and experimental spectra match, 
subtle cancellations in $S^t$ occur among large NME, with residuals modulated by smaller terms. Despite the complexity of the 
full $S^t$ expression at next-to-leading order \cite{Haaranen:2017ovc}, some insights can be gained by elaborating upon  
Eqs.~(\ref{cancel},\,\ref{sNME}), and by recalling that the moments are associated with quadratic forms in both $r$ and $s$. 

The leading NME cancellation operating in $S^t$, as well as in its integral $\mu_0^t$, takes the form of a square of a large linear term in $r$, modulated by a smaller term in $s$. In first approximation, we thus expect the 0$^{\rm th}$ moment to take the form $\mu_0^t \simeq (a-br-cs)^2+d$, with $a$ and $b$ much larger than $c$ and $d$. 
In the plane charted by the $(r,\,s)$ parameters, the equation $\mu_0^t=\mu_0^e$ is then solved by two parallel straight lines with a small slope $-c/b$. In reality, due to subleading NME terms, the perfect square is altered into a more general quadratic form in $(r,\,s)$. Correspondingly, the two straight lines are just the degenerate limit of a conic section, which is actually 
an elongated and slanted ellipse.

\begin{figure}[t!]
\begin{minipage}[c]{0.86\textwidth}
\includegraphics[width=0.21\textwidth]{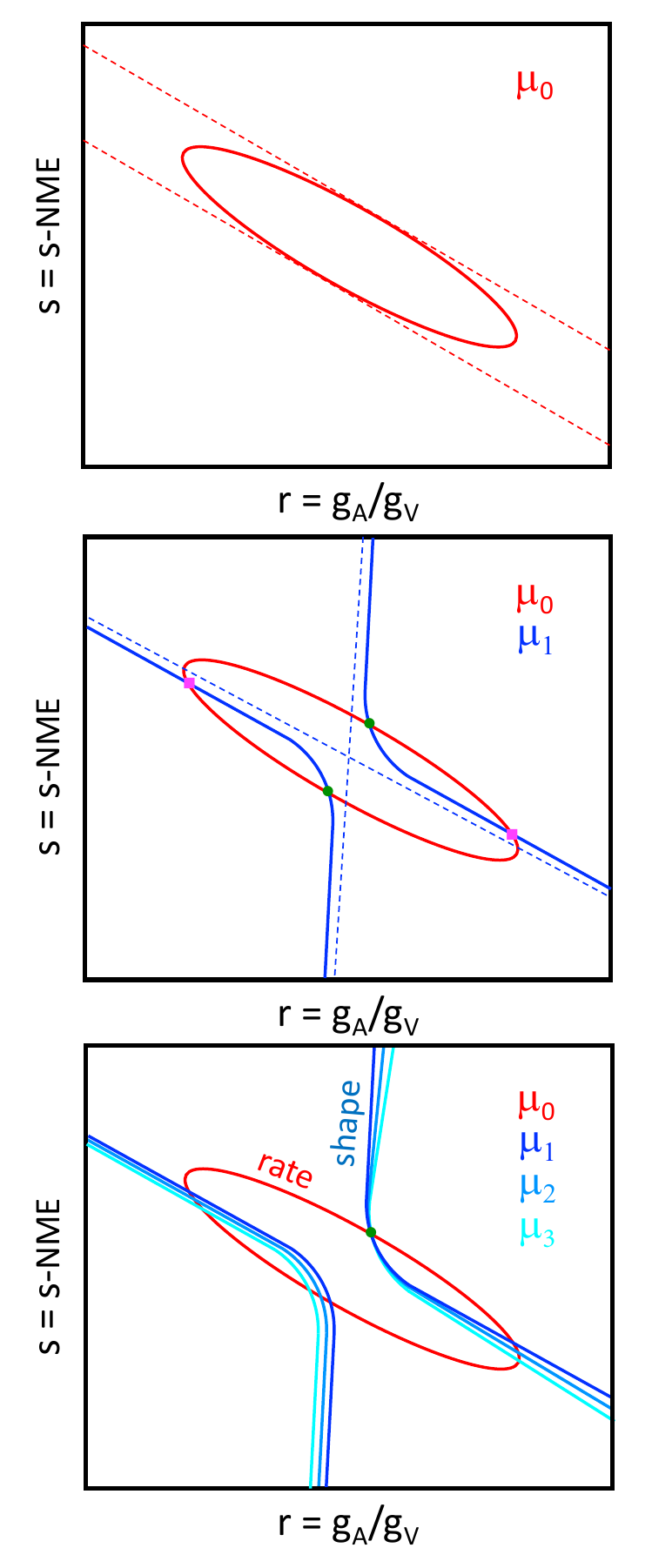}
\vspace*{-3mm}
\caption{\label{Fig_05}
\footnotesize Qualitative expectations for the conic sections obtained by equating the theoretical and experimental moments
in the $(r,\,s)$ plane (in arbitrary scales).
Upper panel: ellipse associated to $\mu_0$, together with the degenerate limit (straight parallel lines). 
Middle panel: hyperbola associated to $\mu_1$, together with it asymptotes (roughly aligned along the vertical axis and the ellipse major axis); the hyperbola crosses the ellipse in four points (solutions). Lower panel: bundle of hyperbolae associated to $\mu_1$ and higher moments, crossing each other and the $\mu_0$ ellipse in a single point (marked with a dot).} \end{minipage}
\end{figure}

Figure~\ref{Fig_05} (upper panel) reports graphically the above qualitative considerations in the plane charted by the $r$ and $s$ parameters.  
The elongated solid ellipse is the locus of points where $\mu_0^t=\mu_0^e$, i.e., where the normalization (event rate above threshold) of the experimental and theoretical spectra match, irrespective of their shapes that may be quite different. The dashed straight lines are a degenerate approximation of the ellipse. 

While the spectrum normalization is associated to $\mu_0$, the spectrum shape is associated to higher moments, starting from $\mu_1$.
Considering that $\mu_1$ is a ratio of quadratic forms, the constraint $\mu^t_1(r,\,s)=\mu^e_1$ leads to 
a quadratic equation as well, identifying another conic section that turns out to be a hyperbola. The appearance of a hyperbola
for $\mu_1$ can be qualitatively understood as follows.

Since the denominator and numerator of $\mu_1$ differ only by a $w_e$ integrand of $O(1)$ in the latter,    
[see Eq.~(\ref{moment})], the associated quadratic forms turn out to have nearly 
proportional coefficients (as confirmed by numerical inspection of Tab.~\ref{Tab1}), 
corresponding to geometrically similar ellipses upon rescaling. The scaling factor 
turns out to be slightly more pronounced along the abscissa $r$, that governs the main NME cancellation term. 
Roughly speaking, the equation $\mu^t_1(r,\,s)=\mu^e_1$ imposes that the ratio between two very similar elliptic forms
(slightly differing along the abscissa) is close to unity. If one form is written as $x^2+y^2-2\rho xy$
(where $x$ and $y$ are generic coordinates), the other is thus obtained by scaling $x$ as $(1+\delta)x$, where $\delta \ll 1$. It is easy to check that,  at first order in $\delta$, the ratio of these two forms is unity for either $x=0$ (corresponding to a vertical line) or for $x=\rho y$ (a slanted line, roughly along the ellipse major axis). These two lines define a degenerate hyperbola, coincident with its two asymptotes: a vertical one and a slanted one. In general, the equation $\mu^t_1(r,\,s)=\mu^e_1$ entails a less simplistic situation:  the hyperbola defined by this equation is not exactly degenerate, and its asymptotes may be slightly tilted with respect to the above expectations.              

Figure~\ref{Fig_05} (middle panel) reports graphically the typical locus of points where $\mu_1^t(r,\,s)=\mu_1^e$, i.e., where the 
average energies of the experimental and theoretical spectra do match. The locus is a hyperbola (blue solid curve), whose branches are relatively close to the degenerate limit (asymptotes, dashed lines). The hyperbola and the ellipse will cross 
in four points, i.e., there will be four different $(r,s)$ solutions to the coupled equations $\mu^t_{0}=\mu^e_{0}$ and $\mu^t_{1}=\mu^e_{1}$. In general, the extreme solutions in $r$ (red dots) correspond to unphysically low or high values of $g_{\rm A}$, and can be discarded a priori; the remaining two solutions (blue dots) typically correspond to negative and positive values of $s$.

With a reasoning similar to $\mu_1$, also the conic sections defined by the second or higher moments are expected to be hyperbolae. 
One then gets an ellipse for $n=0$ (related to the event rate) and a bundle of hyperbolas for $n\geq 1$ 
(related to the spectrum shape).
If a perfect match between theory and data can be achieved, all these conic sections must 
intersect in just one of the four previous points, and diverge to some extent in the others. This case is shown in the lower panel of 
Fig.~\ref{Fig_05}, assuming the solution marked by a dot. In reality, the theory-data match is never perfect, and
the intersections will be slightly separated even in the best possible case.

The above discussion allows to understand some features of previous results reported in \cite{Kostensalo:2020gha}
by using the so-called revised SSM. This approach conflates 
the half-life and spectrum-shape methods (previously conflicting when using only $r$
as a free parameter, see e.g., \cite{Haaranen:2017ovc}), by treating $s$ as an additional degree of freedom.
In particular, let us consider Fig.~2 of \cite{Kostensalo:2020gha}, showing the results of  
fits to an independent $^{113}$Cd 
$\beta$-decay data set \cite{COBRA:2018blx} in the plane charted by $g_{\rm A}$ and the $s$-NMA, for the same three nuclear models 
considered in our work. In that figure,  
while the half-life fit leads to an elliptical solution (akin to the ellipse determined by $\mu_0$ in our
formalism), the spectrum shape fit leads to a band that crosses the ellipse. We can interpret 
such a band as the path taken by the fit algorithm in \cite{Kostensalo:2020gha} to follow (and to fuzzily jump across) two half-branches of the bundle of higher-moment hyperbolae, the other two halves (leading to unphysical values of $g_{\rm A}$) being discarded by construction (the fit was also restricted to $s>0$ therein).  With the spectral moment method discussed herein, the  fit results in
\cite{Kostensalo:2020gha} can be understood in a unified picture.

\end{document}